\documentclass{llncs}
\usepackage{graphicx}
\usepackage{caption}
\usepackage{subcaption}
\usepackage{amssymb}
\usepackage{amsmath}
\usepackage{amsfonts}
\usepackage[english]{babel}
\usepackage[bottom]{footmisc}
\usepackage{wrapfig}
\usepackage{algorithm}
\usepackage[noend]{algpseudocode}
\usepackage[bottom]{footmisc}
\usepackage{algorithm}
\usepackage{algpseudocode}
\usepackage[colorinlistoftodos,prependcaption,textsize=tiny]{todonotes}
\begin{document}

\renewcommand{\qed}{\hfill\blacksquare}

\newtheorem{assumption}{Assumption}

\newtheorem{mydef}{Definition}
\newtheorem{myprob}{Problem}
\newtheorem{observation}{Observation}

\title{Experimental Biological Protocols with Formal Semantics}

\author{Alessandro Abate\inst{2} \and Luca Cardelli\inst{1,2} \and Marta Kwiatkowska\inst{2}  \and Luca Laurenti\inst{2} \and Boyan Yordanov\inst{1}}

\institute{Microsoft Research Cambridge
\and Department of Computer Science, University of Oxford  }

\vspace{-3em}
\maketitle
\begin{abstract}
\vspace{-2em}
Both experimental and computational biology is becoming increasingly automated. Laboratory experiments are now performed automatically on high-throughput machinery, while computational models are synthesized or inferred automatically from data. However, integration between automated tasks in the process of biological discovery is still lacking, largely due to incompatible or missing formal representations. While theories are expressed formally as computational models, existing languages for encoding and automating experimental protocols often lack formal semantics. This makes it challenging to extract novel understanding by identifying when theory and experimental evidence disagree due to errors in the models or the protocols used to validate them. To address this, we formalize the syntax of a core protocol language, which provides a unified description for the models of biochemical systems being experimented on, together with the discrete events representing the liquid-handling steps of biological protocols.
We present both a deterministic and a stochastic semantics to this language, both defined in terms of hybrid processes.  In particular, the stochastic semantics captures uncertainties in equipment tolerances,  making it a suitable tool for both experimental and computational biologists. We illustrate how the proposed protocol language can be used for automated verification and synthesis of laboratory experiments on case studies from the fields of chemistry and molecular programming.
\end{abstract}
\vspace{-3em}
\section{Introduction}
\vspace{-0.7em}
The classical cycle of observation, hypothesis formulation, experimentation, and falsification, which has driven scientific and technical progress since the scientific revolution, is lately becoming automated in all its separate components. Data gathering is conducted by high-throughput machinery. Models are automatically synthesized, at least in part, from data \cite{cardelli2017syntax,dalchau2015synthesizing,andreychenko2011parameter}. Experiments are selected to maximize knowledge acquisition. Laboratory protocols are run under reproducible and auditable software control. However, integration between these automated components is lacking. Theories are not placed in the same formal context as the (coded) protocols that are supposed to test them. Theories talk about changes in physical quantities, while protocols talk about steps carried out by machines: neither knows about the other, although they both try to describe the same process. The consequence is that often it is hard to tell what happened when experiments and models do not match: was it an error in the model, or an error in the protocol? Often both the model and the protocol have unknown parameters: do we use the experimental data to fit the model or to fit the protocol? When most activities are automated, we need a way to answer those questions that is equally automated.

In this paper we present a novel language to model experimental bio-chemical protocols that gives an integrated description of the protocol and of the underlying molecular process.  
From this integrated representation both the model of a phenomenon (for possibly automated mathematical analysis), and the steps carried out to test it (for automated execution by lab equipment) can be separately extracted. This is essential to perform automated model synthesis and falsification by taking also into account uncertainties in both the model structure and equipment tolerances.
\begin{figure}
  \begin{center}
\includegraphics[width=1.1\linewidth]{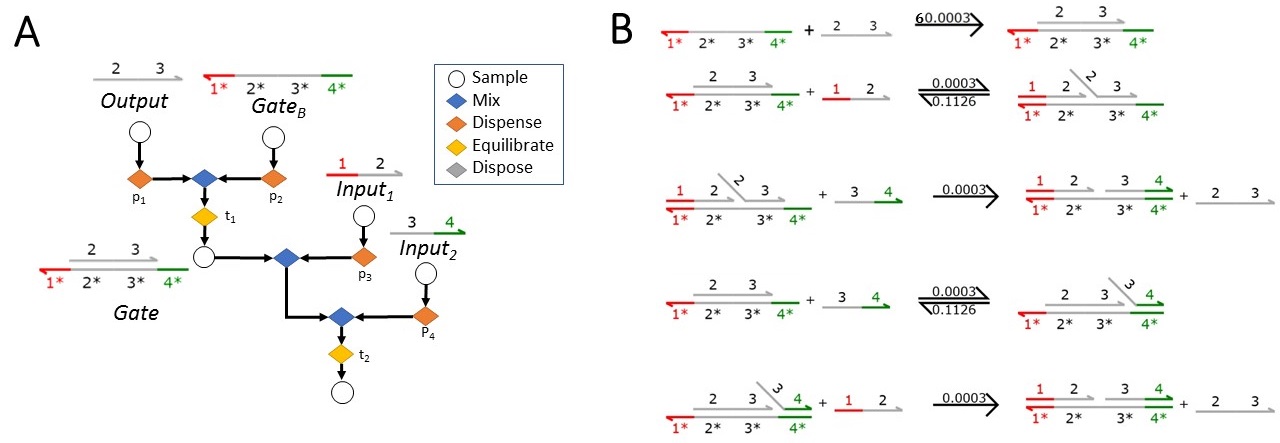}
  \end{center}
  \caption{ (A) Graphical representation of the protocol. Dispose operations discard a sample and are implicitly considered inside Dispense operations (See Section \ref{Sec-CaseStudy} for details) ({B}) Graphical representation of the Chemical Reaction Network (CRN) between the different DNA strands in the considered solution. For example, in the second reaction, strand $\{1^*\}[2$ $3]\{4^*\}$ reacts with $<1^*\, 2>$ at a rate $0.0003$, and there exists an inverse reaction with rate $0.1126$.\vspace{-1.5em}}
    \label{fig:DSDExampleFigure}
\end{figure}
Our goal in this paper is to define a simple core language and focusing on formalizing its semantics. We then show how our language can easily be extended to collect observations of the process and to model complicate protocols. 
 An example of an experimental biological protocol is shown in Example \ref{Example-DSD}.
\begin{example}\label{Example-DSD}
We consider an experimental protocol for DNA strand displacement. DNA strand displacement (DSD) is a design paradigm for DNA nano-devices \cite{Chen2013}. In such a paradigm, single-stranded DNA acts as
signals and double-stranded (or more complex) DNA structures act as gates. The interactions between signals and gates allow one to generate computational mechanisms
that can operate autonomously at the molecular level \cite{seelig2006enzyme}. The DSD programming language has been developed as a means of formally programming and analyzing such devices \cite{lakin2011visual,Chen2013}. In Figure \ref{fig:DSDExampleFigure}, we consider an $AND$ circuit implemented in DSD, 
which can be represented with the Chemical Reaction Netowrk (CRN)  in Figure \ref{fig:DSDExampleFigure}b.  
Strands $Input_1=<1^*$ $2>$ and  $Input_2=<3$ $4^*>$ represent the two inputs, 
while strand $Output=<2$  $3>$ is the output. 
Strand $Gate=\{1^*\}[2$ $3]\{4^*\}$ is an auxiliary strand. 
The $Output$ strand is released only if both the inputs react with the $Gate$ gate.
The protocol in Figure \ref{fig:DSDExampleFigure}a proceeds as follow: $Output$ and $Gate_B$ strands are dispensed from the original samples. Then, they are let evolve for $t_1$ seconds to create $Gate$ strands. Then, the two inputs are dispensed from their samples. The resulting samples are mixed and the resulting solution evolves for $t_2$ seconds. Finally, we collect the final sample and observe the results.
\end{example}


\begin{wrapfigure}{r}{0.5\textwidth}
\vspace{-4em}
  \begin{center}
\includegraphics[width=0.45\textwidth]{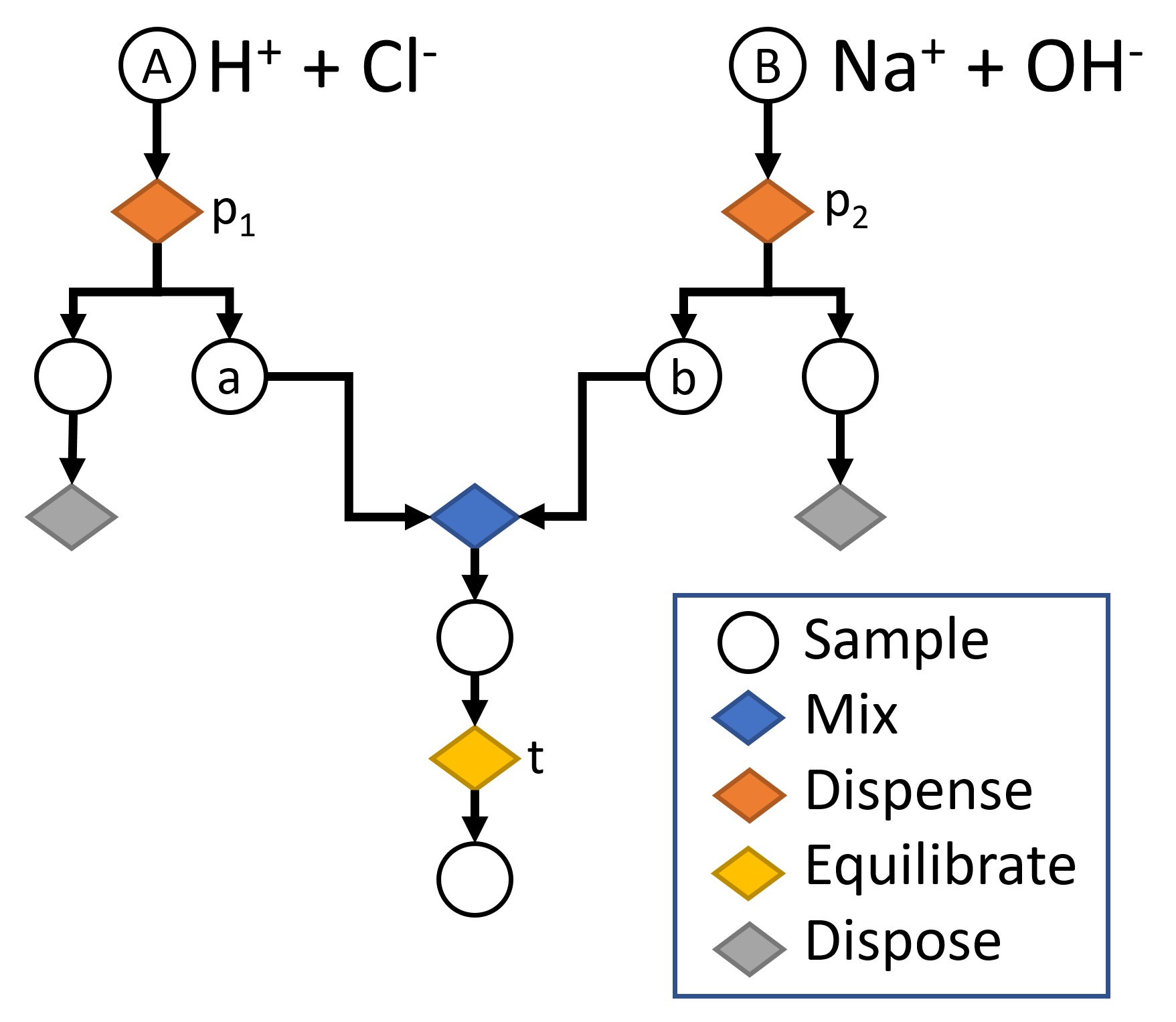}
  \end{center}
  \caption{Graphical representation of an acid-base titration protocol.
  The protocol is initialized with samples A (containing $H^+$ and $Cl^-$) and B (containing $Na^+$ and $OH^-$). Some fraction of each sample ($p_1$ and $p_2$) is mixed together and the resulting sample is let to equilibrate for $t$ seconds.
 }
    \label{fig:TitrationExampleFigure}
    \vspace{-2em}
\end{wrapfigure}
We present two semantics for the introduced language: a deterministic semantics and a stochastic semantics. In both cases, the resulting mathematical model is an hybrid system, where the discrete dynamics are used to map the discrete operations of a lab protocol, while the continuous dynamics model the evolution of the physical variables. In the deterministic semantics, physical variables are modeled in terms of \emph{ordinary differential equations (ODEs)} given by the \emph{rate equations} \cite{ethier2009markov}, while discrete operations are mapped into discrete events that are triggered by some deterministic guards. The stochastic semantics extends the deterministic semantics: it allows one to model uncertainties that are intrinsic in the discrete operations of the protocol, such as those due to lab equipment and whose error ranges have also been standardized (standards ISO 17025 and 8655). Thus, in the resulting stochastic model, the time at which a discrete event happens, may be a random variable with exponential distribution.
We show that the resulting stochastic semantics is a \emph{Piecewise Deterministic Markov Process} (PDMP). That is, a class of Markov stochastic hybrid processes where the continuous variables evolve according to ODEs and the discrete variables evolve by means of random jumps \cite{davis1984piecewise}.  

On examples from chemistry and molecular programming, we demonstrate how our integrated representation allows one  to perform analysis and synthesis of both the discrete steps of the protocol and of the underlying biological system. 
\subsubsection{Related Work}
Several factors contribute to the growing need for a formalization of experimental protocols in biology. First, better record-keeping of experimental operations is recognized as a step towards tackling the ‘reproducibility crisis’ in biology \cite{Freedman2015}. Second, the emergence of ‘cloud labs’ \cite{Hayden2014} 
creates a need for precise, machine-readable descriptions of the experimental steps to be executed.
To address these needs, frameworks allowing protocols to be recorded, shared, and reproduced locally or in a remote lab have been proposed. These frameworks introduce different programming languages for experimental protocols including BioCoder \cite{Ananthanarayanan2010}, Autoprotocol, and Antha \cite{Sadowski2017}. These languages provide expressive, high-level protocol descriptions but consider each experimental sample as a labelled ‘black-box’. This makes it challenging to study a protocol together with the biochemical systems it manipulates in a common framework.

In contrast, we consider a simpler set of protocol operations but capture the details of experimental samples, enabling us to track properties of chemical species (e.g. amounts, concentrations, etc.) as they react during the execution of a protocol. This allows us to formalize and verify requirements for the correct execution of a protocol or to optimize various protocol or system parameters to satisfy these specifications.


\vspace{-1.em}
\section{A Language for Experimental Biological Protocols}
\label{Language}
\vspace{-0.5em}
We introduce the syntax of a new language for modelling experimental protocols. A formal semantics of the language, based on denotational semantics \cite{scott1971toward}, is then discussed. The physical process underlying a biological experimental protocol is modeled  as a Chemical Reaction Systems (CRS). A CRS is a Chemical Reaction Network (CRN) with a given initial condition and will be formally defined in the next section (Definition \ref{Definition-CRS}).
\begin{mydef}{(Syntax of a Protocol)}\label{LangSynt2} Given a set of variables $Var$, the syntax of a protocol $P$ for a given fixed CRN $\mathcal{C}=(\mathcal{A},\mathcal{R})$ is 
\begin{align*}
P=\quad \quad \quad\quad \quad \quad\quad\quad \quad \quad \quad \quad \quad x   \quad &\text{(sample variable)}\\
 (x_0,V,T)\quad  & (\text{initial condition})\\
 Mix(P_1,P_2) \quad & (\text{mix samples})\\
 let\, x=P_1 \, in \, P_2 \quad & \text{(define variable)} \\
let \, x,y=Dispense(P_1,p)\, in\, P_2\quad & \text{(dispense samples)}\\
 Equilibrate(P,t) \quad & (\text{let time pass})\\
 Dispose(P) \quad & (\text{discard P})
\end{align*}
where $T,V,t\in \mathbb{R}_{\geq 0},x_0\in \mathbb{R}^{|\mathcal{A}|}, x,y \in Var$, $p\in \mathbb{R}_{(0,1)}$. Moreover, let-bound variables must occur (as free variables) exactly once in $P_2$.
\end{mydef}
\noindent

A protocol $P$ yields a sample, which is the result of operations of Equilibrate, Mix, Dispose  and Dispense, over a CRS. This syntax allows one to create and manipulate new samples using Mix (put together different samples), Dispense (separate samples) and Dispose (discard samples) operations.
Note that the CRN is common for all samples. However, different samples may have different initial conditions. The single-occurrence (linearity) restriction implies that a sample cannot be duplicated or eliminated from the pool.

\begin{example}
\label{IllustrativeExample}
We use $let\,x,\_ = Dispense(P_1, p)\,in\,P_2$ as a short-hand for $let\,x,y = Dispense(P_1,p)\,in\,Mix(Dispose(y),x)$.
Given a CRN $\mathcal{C}=(\{H^+,Cl^-,Na^+,$ $OH^-,H_2O \},\mathcal{R})$, where $\mathcal{R}=\{Na^{+} + OH^{-} + H^{+} + Cl^{-} \to^{k} H_2O + Na^{+} + Cl^{-} \}$,
the protocol (call it $Pro_1$) represented graphically in Figure~\ref{fig:TitrationExampleFigure} is defined formally as
\begin{align*}
Pro_1 = &let\,A = ([(H^+, 0.1 M); (Cl^-, 0.1 M)], 1.0 mL, 298.15 K) \,in\\
&let\,B =  ([(Na^+, 0.1 M); (OH^-, 0.1 M)], 1.0 mL, 298.15 K) \,in\\
&let\,a,\_ = Dispense(A,p_1)\,in\\
&let\,b,\_ = Dispense(B,p_2)\,in\\
&Equilibrate(Mix(a,b), t).
\end{align*}
{In the formula above, $[(H^+, 0.1 M); (Cl^-, 0.1 M)]$ is a short-hand for vector $[0.1,0.1,0,0,0]$ representing the initial concentration of the species in sample $A$ for CRN $\mathcal{C}$, where we made clear that the concentration is specified in molar units (M).}
\end{example}
The following equivalences can be shown structurally, namely based on the definitions of substitution and free-variables reported in the Appendix (Definitions \ref{DefinitionSubstitution} and \ref{Def-FreeVariables}). 
\begin{proposition}{(Equivalence Relationships)}
\begin{align*}
    &let \, x=P_1 \, in \, P_2 \,\,=\,\,P_2\{ x\leftarrow P_1\}\\
    &let \, x=P_1 \, in \, P_2 \,\,=\,\,let \, y=P_1 \, in \, (P_2\{ x\leftarrow y\}) \text{ for $y \not \in FV(P_2)$}, 
\end{align*}
where $P_2\{ x\leftarrow P_1\}$ is the is the capture-avoiding substitution of $P_2$ for $x$ in $P_1$, and $FV(P_2)$ are the free variables of $P_2$, as defined in Definitions \ref{DefinitionSubstitution} and \ref{Def-FreeVariables} in the Appendix. 
\end{proposition}
\noindent
{We stress that in order to define a semantics for the protocol language in Definition \ref{LangSynt2}, we require a pair $(P,\mathcal{C})$, where $P$ is a protocol and $\mathcal{C}$ is a CRN. In the next Section we formally introduce CRNs. However, we should also stress that many languages exist to represent CRNs. For instance, graphical languages or implicit representations, as those that we use in Example \ref{fig:DSDExampleFigure}, where the set of reactions can be determined just from the structure of the initial DNA strands, by the rules of DNA strand displacement \cite{phillips2009programming}. In this paper, we do not require a particular representation language for CRNs. We simply assume that we can always extract a representation of a CRN, which matches the definition given in the next Section. }

\vspace{-1.em}
\section{Chemical Reaction Networks}
\label{section-CRN}
A CRN $\mathcal{C}=(\mathcal{A},\mathcal{R})$ is a pair of finite sets, where $\mathcal{A}$ denotes a set of \emph{chemical species}, $|\mathcal{A}|$ is its cardinality, and $\mathcal{R}$ denotes a set of reactions. 
A \emph{reaction} $\tau \in \mathcal{R}$ is a triple $\tau=(r_{\tau},p_{\tau},k_{\tau})$, where $r_{\tau} \in  \mathbb{N}^{|\Lambda|}$ is the \emph{source complex}, $p_{\tau} \in  \mathbb{N}^{|\Lambda|}$ is the \emph{product complex} and $k_{\tau} \in \mathbb{R}_{>0} $ is the coefficient associated with the rate of the reaction.
The quantities $r_{\tau}$ and $p_{\tau}$  represent the stoichiometry of reactants and products.
Given a reaction $\tau_1=(  [1,0,1],[0,2,0],k_1 )$, we often refer to it visually as $\tau_1 : \lambda_1 + \lambda_3 \, \rightarrow^{k_1}  \,    2\lambda_2 $.
The \emph{net change} associated to $\tau$ is defined by $\upsilon_{\tau}=p_{\tau} - r_{\tau}$. 

Many models have been introduced to study CRNs \cite{bortolussi2016approximation,cardelliprogramming,ethier2009markov,cardelli2016stochastic}. Here we consider the {\it rate equations} \cite{ethier2009markov}, which describe the time evolution of the concentration of the species in $\mathcal{C}$, in a sample of temperature $T$ and volume $V$, as follows:  
 \begin{equation}\label{eq:ODE_mean}
 \frac{d \Phi(t)}{dt}=F(t)=\sum_{\tau \in \mathcal{R}}\upsilon_{\tau}\cdot  \gamma_{S}(\Phi(t),k_{\tau},V,T), 
 \end{equation}
where $\gamma_S(\Phi(t),k_{\tau},V,T))$ is the propensity rate, and in case of mass action kinetics we have
$$\gamma_{S}(\Phi(t),k_{\tau},V,T))= k_\tau(T) \prod_{S \in \Lambda}\Phi_{S}^{r_{S,\tau}}(t),$$  where
$\Phi_{S}$ and $r_{S,\tau}$ are the components of vectors
$\Phi$ and $r_{\tau}$
relative to species $S$, and where in $k_\tau(T)$ we make explicit the dependence from temperature $T$. 
\begin{mydef}(Chemical Reaction System)\label{Definition-CRS}
A  \emph{chemical reaction system} (CRS) $C=(\mathcal{A},\mathcal {R},x_0)$ is defined as a tuple, where $(\mathcal{A},\mathcal{R})$ is a CRN and $x_0 \in \mathbb{N}^{|\Lambda|}$ represents its initial condition.
\end{mydef}

\begin{example}
\label{ReactionExampleHCL}
Consider the CRS $C=(\mathcal{A},\mathcal{R},x_0),$ evolving in a volume $V$ and at  temperature $T$, where $\mathcal{A}=\{H_2O,Na^+,OH^-,Cl^-,H^+ \}$ and $\mathcal{R}$ is composed of the following reactions: 
\begin{align*}
&Na^{+} + OH^{-} + H^{+} + Cl^{-} \to^{k} H_2O + Na^{+} + Cl^{-}
\end{align*}
where $k=2.81e^{-10}$ is the rate at temperature $T=298$ Kelvin.
Then, according to Equation  \eqref{eq:ODE_mean}, we have that the state of $H^+$ is given by the solution of the following ordinary differential equation:   
\begin{align*}
    \frac{d H^+(t)}{dt}=& -k Na^{+}(t)OH^{-}(t) H^{+}(t) Cl^{-}(t), 
\end{align*}
with $H^+(0)=\frac{x_{0,H^+}}{V},$ where $x_{0,H^+}$ is the component of $x_0$ corresponding to $H^+$.
\end{example}

In order to introduce a formal semantics for experimental protocols, we first need to define a formal semantics for a CRS, which has been only  introduced informally in the previous section. 
Let $S=(\mathbb{R}^{|\mathcal{A}|}\times\mathbb{R}_{\geq 0}\times\mathbb{R}_{\geq 0})$ be a sample. Then, we define the semantics for a CRS as follows.  
\begin{mydef}(CRS Semantics)
\label{CRSDef}
Let $\mathcal{C}=(\mathcal{A},\mathcal{R})$ be a CRN, $x_0 \in \mathbb{R}^{|\mathcal{A}|}_{\geq 0},$ $V,T\in \mathbb{R}_{\geq 0}$ be the initial concentration (moles), volume (liters) and temperature (degrees Kelvin). Call $F(V,T):\mathbb{R}^{|\mathcal{A}|}\to \mathbb{R}^{|\mathcal{A}|}$ the drift at volume $V$ and temperature $T$ for $\mathcal{C}$. Then, the semantics of the CRS $(\mathcal{A},\mathcal{R},x_0)$ at volume  $V$, temperature $T$ and time $t$, for a time horizon $H\in \mathbb{R}_{\geq 0}\cup\{\infty \}$,$$[\![\cdot]\!]:(CRS\times \mathbb{R}_{\geq 0} \times \mathbb{R}_{\geq 0})\to \mathbb{R}_{\geq 0}\cup\{\infty\}\to \mathbb{R}_{\geq 0}\to S$$  is defined as 
\begin{align*}
[\![ ((\mathcal{A},&\mathcal{R},x_0),V,T)]\!](H)(t)=\\
&let\, G:[0...H)\to \mathbb{R}^{|\mathcal{A}|} \text{ be the solution of } G(t')=x_0+\int_{0}^{t'} F(V,T)(G(s))ds\\
&(G(t),V,T),
\end{align*} 
{where the above operation reads as follows: `first line' = `third line', where $G$ is defined as in `second line'.}
If for such an $H$, $G$ is not unique, then we say that $[\![ ((\mathcal{A},\mathcal{R},x_0),V,T)]\!]$ $(H)(t)$ is ill posed.
\end{mydef}
In Definition \ref{CRSDef} we have explicitly introduced a dependence on a time horizon $H$, because it may happen that the solution of the rate equations is defined only for a finite time horizon \cite{ethier2009markov}.

\section{Deterministic Semantics of Experimental Protocols}
In an experimental protocol discrete operations are mixed with physical variables, namely concentration of the species of a CRN that evolve continuously in time.  
We first consider a deterministic semantics for the language presented in Definition \ref{LangSynt2}. 
Then, in the next section, we extend such a semantics in order to take into account errors and inaccuracies within a protocol, which in practice can be quite relevant: this leads to probabilistic semantics. 

The deterministic semantics of a protocol $P$ for a CRN $\mathcal{C}=(\mathcal{A},\mathcal{R})$, 
under a given environment $\rho:Var \to S$, 
is a function $[\![ P]\!]^{\rho}:(Var \to S)\to S$, where $S$ is a sample as defined in Section \ref{section-CRN},  defined inductively as follows. 
\begin{mydef}{(Deterministic Semantics of a Protocol)}\label{DetSem}
Let $S=(\mathbb{R}^{|\mathcal{A}|}\times\mathbb{R}_{\geq 0}\times\mathbb{R}_{\geq 0})$, then the deterministic semantics of a protocol $P$ for CRN $\mathcal{C}=(\mathcal{A},\mathcal{R})$, under environment $\rho:Var \to S$  is defined inductively as follows
\begin{align*}
    &[\![x]\!]^{\rho}=\rho(x)\\
     &[\![x_0,V,T]\!]^{\rho}=(x_0,V,T)\\
     &[\![ Mix(P_1,P_2)]\!]^{\rho}=\\
     &\quad let\, (x_0^1,V_1,T_1)= [\![P_1]\!]^{\rho}\\
     &\quad let\, (x_0^2,V_2,T_2)= [\![P_2]\!]^{\rho}\\
     &\quad (\frac{x_0^1V_1 +x_0^2V_2  }{V_1+V_2},V_1+V_2,\frac{T_1V_1 +T_2V_2  }{V_1+V_2} )\\
     &[\![let \, x=P_1\, in\, P_2]\!]^{\rho}=\\
      &\quad let \,(x_0,V,T)= [\![P_1]\!]^{\rho}\\
      &\quad let \,{\rho_1}=\rho\{x \leftarrow (x_0,V,T)\}\\
      &\quad [\![P_2]\!]^{\rho_1}
      \end{align*}
      \begin{align*}
      &[\![let \, x,y=Dispense(P_1,p)\, in\, P_2]\!]^{\rho}=\\
      &\quad let \,(x_0,V,T)= [\![P_1]\!]^{\rho}\\
      &\quad let \,{\rho_1}=\rho\{x \leftarrow (x_0,V\cdot p,T),y \leftarrow (x_0,V\cdot (1-p),T)\}\\
      &\quad [\![P_2]\!]^{\rho_1}\\
       &[\![ Equilibrate(P,t)]\!]^{\rho}=\\
       &\quad let\, (x_0,V,T)=[\![P]\!]^{\rho}\\
      &\quad  [\![(\mathcal{A},\mathcal{R},x_0),V,T)]\!](H)(t)\\
      &[\![Dispose(P)]\!]^{\rho}=(0^{|\Lambda|},0,0), 
\end{align*} 
where $H\in \mathbb{R}_{\geq 0}$ is such that for any $Equilibrate(P,t)$, 
$[\![(\mathcal{A},\mathcal{R}),$ $x_0,V,T)]\!](H)(t)$ is well posed. 
If such an $H$ does not exist, we say that $P$ is ill posed.
\end{mydef}
The above semantics identifies a protocol with the concentration of the species, the volume,  and the temperature of the sample at final time.

\begin{example}
Consider the protocol $Pro_1$ introduced in Example \ref{IllustrativeExample}. 
The CRN of the system comprises the reactions given in the CRN in Example \ref{ReactionExampleHCL}.
According to Definition \ref{DetSem}, 
the state of variable  $H^+$ at the end of $Pro_1$ is given by the solution of the following equation:
\begin{align*}
   [\![Pro_1]\!]^{\rho}(H^+)=H^+(0)& -\int_0^{t } k Na^{+}(s)OH^{-}(s) H^{+}(s) Cl^{-}(s)ds, 
\end{align*}
where $ H^+(0)= \frac{p_1 0.1+p_2 10^{-7.4} }{p_1+p_2}, $ $\rho$ is any environment, and $[\![Pro_1]\!]^{\rho}(H^+)$ stands for the component relative to $H^+$ of the sample after the execution of the protocol. 
\end{example}
\vspace{-0.5em}
\section{Stochastic Semantics of Experimental Protocols}
\vspace{-0.5em}
In this Section we introduce the stochastic semantics for an experimental protocol, and show that any program defined according to Definition \ref{LangSynt2} can be mapped onto a Piecewise Deterministic Markov Processes (PDMPs) \cite{davis1984piecewise}. 
PDMPs, introduced in Section \ref{Section-PDMP}, are a class of stochastic hybrid systems where continuous variables evolve deterministically according to a system of ordinary differential equations (ODEs), while discrete operations may be probabilistic, and introduce noise in the system.
\vspace{-0.5em}
\subsection{Piecewise Deterministic Markov Process}\label{Section-PDMP}
\vspace{-0.5em}
The syntax of a PDMP is given as follows. 
\begin{mydef}
\label{PPDMP}
A Piecewise Deterministic Markov Process (PDMP) $\mathcal{H}$ is a tuple $\mathcal{H}=(\mathcal{Q},d,\mathcal{G},F,\Lambda,R)$, where
\begin{itemize}
    \item $\mathcal{Q}=\{q_1,...,q_{|\mathcal{Q}|}\}$ is the set of \emph{discrete modes}
    \item $d \in \mathbb{N}$ is the dimension of the state space of the \emph{continuous dynamics}. The hybrid state space is defined as $\mathcal{S}=\cup_{q\in \mathcal{Q}}\{ q\}\times \mathbb{R}^{d}$
    \item $\mathcal{G}: \mathcal{Q}\times \mathbb{R}^{d}\to \{0,1 \}$ is a set of guards 
    \item $F:\mathcal{Q}\times \mathbb{R}^{d} \to \mathbb{R}^{d}$ is a family of vector fields 
    \item $\Lambda:\mathcal{S}\times \mathcal{Q}\to \mathbb{R}_{\geq 0}$ is an \emph{intensity function}, where for  $(q_i,x)\in \mathcal{S},q_j\in \mathcal{Q}$, we define $\Lambda((q_i,x),q_j)=\lambda_{i,j}(x)$ and $\lambda_{q_i}(x)=\sum_{q_j\neq q_i}\lambda_{i,j}(x)$
    \item $R: \mathcal{B}(\mathcal{S})\times \mathcal{S}\to [0.1]$ is the \emph{reset} function, which assigns to each $(q,x)\in \mathcal{S}$ a probability measure $R(\cdot,q,x)$ on $(\mathcal{S},\mathcal{B}(\mathcal{S}))$, where $\mathcal{B}(\mathcal{S})$ is the smallest $\sigma-$algebra on $\mathcal{S}$ containing all the sets of the form $\cup_{q\in\mathcal{Q}}\{q\}\times A_q,$ where $A_q$ is a measurable subeset of $\mathbb{R}^{d}$.
    \end{itemize}
\end{mydef}
 For ${t}\in\mathbb{R}_{\geq 0},q\in \mathcal{Q},x\in\mathbb{R}^{d}$, we call $\Phi(q,t,x)$ the solution of the following differential equation:  
$$ \frac{d \Phi(q,t,x)}{dt}=F(q,\Phi(q,t,x)),  \quad \Phi(q,0,x)=x.$$
The solution of a PDMP is a stochastic process $Y=(\alpha,X),$ whose semantics is classically defined according to the notion of execution (see Definition \ref{PDMPExecution} below) \cite{davis1993markov}. In order to introduce such a notion, we define the exit time $t^*(q,x,\mathcal{G})$ as
\begin{align}
    \label{hitting time}
    t^*(q,x,\mathcal{G})=\inf\{t\in \mathbb{R}_{\geq 0}\, | \, \mathcal{G}(q,\Phi(q,t,x))=1\}
\end{align}
and the  \emph{survival function}
$
    f(q,t,x)=\begin{cases}
    e^{-\int_0^t\lambda_{q}(\Phi(q,\tau,x))d \tau}      & \quad \text{if } t<t^*(q,x,\mathcal{G})\\\
    0  & \quad \text{otherwise.}\\
  \end{cases}.$
Here $t^*(q,x,\mathcal{G})$ represents the first time instant, starting from state $(q,x)$, when the guard set is reached by a solution of the process; further $f(q,t,x)$ denotes the probability that the system remains within $q$, starting from $x$, at time $t$ \cite{davis1984piecewise}, which depends on random arrivals induced by the intensity function $\Lambda$. The semantics of a PDMP for initial condition $(q_0,x_0)$ is provided next. 
\vspace{-0.5em}
\begin{mydef}{(Execution of PDMP $\mathcal{H}$)}\label{PDMPExecution}

\textbf{Set} $t:=0$

\textbf{Set} $(\alpha(0),X(0)):=(q_0,x_0)$

\textbf{While} $t<\infty$ 

\quad \textbf{Sample} $\mathbb{R}_{\geq 0}$-valued random variable $T$ such that $$Prob(T>\bar{t})=f(\alpha(t),\bar{t},X(t))$$

\quad$\forall \tau\in [t,t+T)$ \textbf{Set} $(\alpha(\tau),X(\tau)):=(\alpha(t),\Phi(\alpha(t),\tau-t,X(t)))$

\quad  \textbf{If $t+T<\infty$}

\quad  \quad  \textbf{Sample} $(\alpha(t+T),X(t+T))$ according to

\quad \quad \quad \quad $R(\cdot,(\alpha(t),\Phi(\alpha(t),T,X(t)))$

\quad   \textbf{End If}

\quad \textbf{Set} $t:=t+T$

\textbf{End While}
\end{mydef}
For further details on PDMPs and on their measure theoretic properties we refer to \cite{davis1984piecewise}.

\subsection{Stochastic Semantics}
Let us recall that the semantics of Definition \ref{DetSem} are fully deterministic. 
However, it is often the case that operations of $Dispense$ and $Equilibrate$ are stochastic in nature, 
due to the fact that they are performed by humans and in view of experimental inaccuracies related to lab equipment. 
In what follows, we encompass these features  
by extending the semantics, previously defined as deterministic, with stochasticity. 
More precisely, we account for the following: 
\begin{itemize}
    \item in the $Equilibrate(P,t)$ step, time is sampled from a distribution;  
    \item the resulting volume after a $Dispense$ step is sampled from a distribution. 
\end{itemize}
The first characteristic models the fact that in real experiments the system is not equilibrated for exactly $t$ seconds, as it may start or be stopped at different time instants, and it accounts for the fact that after a mix of samples well mixed conditions are not reached instantaneously;  whereas the second feature takes into account the experimental errors associated to pipetting devices whose ranges have been standardized (standard ISO $8655$). 
For the first feature, consider the function $\mathcal{T}(t',t)=e^{-\frac{t'}{t}}$, 
defined for two values $t',t\in \mathbb{R}_{\geq 0}$:  
this corresponds to the density function of an exponential random variable, 
modelling random arrivals. 
For the second feature,   
let $\mathcal{B}(\mathbb{R}_{\geq 0}^m)$ be the Borel sigma-algebra over $\mathbb{R}^m_{\geq 0}$, $m>0.$ Then, we consider the following function $\mathcal{D}:\mathcal{B}(\mathbb{R}_{[0,1]})\times \mathbb{R}_{\geq 0}\times \mathbb{R}_{[0,1]}\to [0,1]$, which assigns  to $\mathcal{D}(\cdot,V,p)$ a probability measure in $\mathcal{B}(\mathbb{R}_{[0,1]})$. 
Function $\mathcal{D}$ is used to reset the volume randomly, after a discrete operation. 
(As an anticipation of upcoming results, 
notice that both functions $\mathcal T$ and $\mathcal D$ can be mapped to elements in the syntax of a PDMP model.)

We define the \emph{Stochastic Semantics} of a protocol as an extension of the deterministic ones from Definition \ref{DetSem}. 
For the sake of compactness, we write explicitly only the operators that differ from the earlier definition. 

\begin{mydef}{(Stochastic Semantics of a Protocol)}\label{DPDMPSem}
Let $S=(\mathbb{R}^{|\mathcal{A}|}\times\mathbb{R}_{\geq 0}\times\mathbb{R}_{\geq 0})$, then the semantics of a protocol $P$ for CRN $\mathcal{C}=(\mathcal{A},\mathcal{R})$, under environment $\rho:Var \to S$ and functions $\mathcal{T}$, $\mathcal{D}$, as defined above, is defined inductively as follows
\begin{align*}
      &[\![let \, x,y=Dispense(P_1,p)\, in\, P_2]\!]^{\rho}=\\
      &\quad let \,(x_0,V,T)= [\![P_1]\!]^{\rho}\\
      &\quad let\, p'\,being\,sampled\,from\,\mathcal{D}(\cdot,V,p)\\
      &\quad let \,{\rho_1}=\rho\{x \leftarrow (x_0,V\cdot p',T),y \leftarrow (x_0,V\cdot (1-p'),T)\}\\
      &\quad [\![P_2]\!]^{\rho_1}\\
       &[\![ Equilibrate(P,t)]\!]^{\rho}=\\
       &\quad let\, (x_0,V,T)=[\![P]\!]^{\rho}\\
       &\quad let\,\mathcal{I}\,be\,a\,\mathbb{R}_{\geq 0}-valued\,\text{random  variable such that for $s\in \mathbb{R}_{\geq 0}$}\\
       &\quad\quad\quad Prob(\mathcal{I}>s)=\mathcal{T}(s,t)\\
      &\quad  [\![(\mathcal{A},\mathcal{R},x_0),V,T)]\!](H)(\mathcal{I}),
\end{align*} 
where $H\in \mathbb{R}_{\geq 0}$ is such that for any $Equilibrate(P,t)$, and any $\mathcal{I}$ random variable such that $Prob(\mathcal{I}>s)=\mathcal{T}(s,t),$ $[\![(\mathcal{A},\mathcal{R}),x_0,V,T)]\!](H)(\mathcal{I})$ is well posed with probability $1$. If such an $H$ does not exist, we say that $[\![P]\!]^{\rho}$ is ill posed. 
\end{mydef}
$\mathcal{D}$ is a transition kernel that depends only on the current state of the system. $\mathcal{T}$ is the cumulative probability distribution of a random variable with exponential distribution.  As a consequence, according to Definition \ref{PDMPExecution}, $[\![P]\!]^{\rho}$ induces semantics that are solution of a PDMP. $\mathcal{T}$ determines the probability of changing discrete state and $\mathcal{D}$ acts as a probabilistic reset,  there are no guards, and the continuous dynamics evolve according to the ODE in Definition \ref{CRSDef}.
More formally, given a protocol $P$ and an environment $\rho$, 
$[\![P]\!]^{\rho}$ induces semantics that correspond to the solution of a PDMP $\mathcal{H}=(\mathcal{Q},d,\mathcal{G},F,\Lambda,R)$ as per Definitions \ref{PPDMP} and \ref{PDMPExecution}.  
In  $\mathcal{H}$, 
$\mathcal{Q}$ represents the set of discrete operations, $d=|\mathcal{A}|+1$ denotes the continuous dimension (the number of continuous variables plus one ODE for modeling the time evolution). 
The vector field $F$ is given by Definition \ref{CRSDef}, 
with an additional clock variable $time$ representing time as  $\frac{d time}{dt}=1$. 
For each $Equilibrate(P,t)$ step, t is sampled from $\mathcal{T}$. $\mathcal{D}$ is a reset associated to Dispense operations.

\medskip

We can now leverage results from the analysis of PDMP models and export them over the protocol language. 
The following assumptions guarantee that $[\![P]\!]^{\rho}$ exists, 
is a strong Markov process, 
and allow us to exclude pathological Zeno behaviours \cite{davis1984piecewise,kouretas2006stochastic}. 
\begin{assumption}
\label{Assumptions} \hfill
\begin{itemize}
    \item Let  $A_0,A_1\subset \mathcal{B}(\mathbb{R}_{[ 0,1]})$ be the smallest sets in $\mathcal{B}(\mathbb{R}_{[ 0,1]})$ containing respectively $0$ and $1$. Then,  $\mathcal{D}(A_{0},V,p)=\mathcal{D}(A_{1},V,p)=0$ for any $p \in (0,1),V \neq 0 .$ That is, the Volume of a sample after a dispense is zero with probability zero.
    \item Let $F$ be the drift term of the rate equations (Eqn \eqref{CRSDef}). Then, $F$ is a globally Lipschitz function.
    \item For any $Equilibrate(\cdot,t)$ we have that $t>0$.
\end{itemize}
\end{assumption} 
Let us interpret these assumptions over the protocol languages.  
The first assumption guarantees that the volume of a non-empty sample is almost-surely not equal to $0$. 
The second assumption guarantees that the solution of \eqref{CRSDef} exists and does not hit infinity in finite time. 
This excludes non-physical reactions like $X + X \to X + X + X$. 
The third assumption guarantees that we have a finite number of jumps over a finite time, thus excluding Zeno behaviours \cite{davis1984piecewise,davis1993markov}.

\begin{example}
Consider the protocol introduced in Example \ref{IllustrativeExample}. 
For $\sigma_1>0,A \subset  \mathbb{R}_{[0.1,0.8]}$. 
Assume that 
$\mathcal{D}(A,p,\bar{V})=\frac{ \int_{A} e^{-\frac{x-p}{2 \sigma_1^2}}dx }{\int_{0.1}^{0.8} e^{-\frac{x-p}{2 \sigma_1^2}}dx}. $
That is, $\mathcal{D}(\cdot,p,V)$ is a truncated Gaussian measure centered at $p$ and independent of the volume. 
Then, according to Definition \ref{DPDMPSem}, we have the following final value for $H^+$:  
\begin{align*}
    H^+(\mathcal{I})=&H^+(0)- \int_{0}^\mathcal{I}k Na^{+}(s)OH^{-}(s) H^{+}(s) Cl^{-}(s)ds, 
\end{align*}
with  $HCl(0)= \frac{V_1 0.1+V_2 10^{-7.4} }{V_1+V_2}. $ 
Here $\mathcal{I}$ is a random variable with an exponential distribution with rate $\frac{1}{T},$ 
$V_1$ is a random variable sampled from $\mathcal{D}(\cdot,p_1,1)$, 
and $V_2$ is a random variable sampled from $\mathcal{D}(\cdot,p_2,1)$. 
\end{example}
\begin{remark}
The deterministic semantics (Definition \ref{DetSem}) can also be mapped into the framework of PDMPs. More specifically, Definition \ref{DetSem} induces a PDMP $\mathcal{H}=(\mathcal{Q},d,\mathcal{G},F,\Lambda,R)$, where $\mathcal{Q}$ is the set of discrete operations, $\Lambda((q_i,x),q_j)=0 $ for any $q_i,q_j \in \mathcal{Q},x \in \mathbb{R}^{d}$, $\mathcal{G}$ is a set of guards hitting the changes in the discrete locations when the variable modelling the time reaches a threshold, $F$ is given by Definition \ref{CRSDef}, and the reset R is a Dirac delta function.  
\end{remark}

\vspace{-1.em}
\section{Extending the Protocol Language with Observations}
\vspace{-0.5em}
 The language introduced in Section \ref{Language} can be extended in a number of directions, according to specific scenarios envisioned for the protocols. 
 For instance, a common laboratory task is to take observations of the state of the samples handled by a protocol. 
 That is, often it is useful to store the state of the system at different times or when a particular event happens. As some of the events may be stochastic, in general it is not possible to know before the simulation starts when a particular event happens. Consequently, observations need to be included in the language.
 \begin{mydef}{(Extended Syntax).}\label{LangSynt2Extended} Given a set of variables $Var,$ the syntax of a protocol for a given fixed CRN $\mathcal{C}=(\mathcal{A},\mathcal{R})$ and $idn \in \mathbb{N}$ is 
\begin{align*}
P=\quad \quad \quad\quad \quad \quad\quad\quad \quad \quad \quad\quad \quad \quad  x   \quad &\text{(sample variable)}\\
 (x_0,V,T)\quad  & (\text{initial condition})\\
 Mix(P_1,P_2) \quad & (\text{mix samples})\\
 let\, x=P_1 \, in \, P_2 \quad & \text{(define variable)} \\
let \, x,y=Dispense(P_1,p)\, in\, P_2\quad & \text{(dispense samples)}\\
 Equilibrate(P,t) \quad & (\text{let time pass})\\
 Dispose(P) \quad & (\text{discard P})\\
  Observe(P,idn) \quad & (\text{observe sample})
\end{align*}
where $T,V,t\in \mathbb{R}_{\geq 0}, x,y \in Var$, $p\in \mathbb{R}_{[0,1]}$. Moreover, let-bound variables must occur exactly once ({ that is, be free}) in $P_2$.
\end{mydef} 
 $Observe(P,idn)$ makes an observation of protocol $P$ after its execution, and identifies such an observation with identifier $idn$. 
 In order to include observations we extend the semantics as detailed next, 
 where we consider in detail just the deterministic semantics, focusing on a few key operators. 
 Extensions to the other operators and to Stochastic Semantics follow intuitively. 
 \begin{mydef}{(Extended Deterministic Semantics)}\label{ExtDetSem}
For CRN $\mathcal{C}=(\mathcal{A},\mathcal{R})$ let $S=\mathbb{R}^{|\mathcal{A}|}\times \mathbb{R}_{\geq 0}\times \mathbb{R}_{\geq 0},$ $Obs=\mathbb{R}^{|\mathcal{A}|}\times \mathbb{N}\times \mathbb{R}_{\geq 0},$ $Obs^*,$ an eventually empty set of $Obs$ and $\mathcal{M}=S\times Obs^* \times \mathbb{R}_{\geq 0} .$ The semantics of a protocol $P$,  under environment $\rho:Var \to \mathcal{M}$, is a function $[\![ P]\!]:(Var \to \mathcal{M})\times \mathbb{R}_{\geq 0}\to \mathcal{M} $ defined inductively as follows
\begin{align*}
     &[\![ Mix(P_1,P_2)]\!]^{\rho}_t=\\
     &\quad let\, ((x_0^1,V_1,T_1),Obs_1,t_1)= [\![P_1]\!]^{\rho}_{t}\\
     &\quad let\, ((x_0^2,V_2,T_2),Obs_2,t_2)= [\![P_2]\!]^{\rho}_{t}\\
     &\quad( (\frac{x_0^1V_1 +x_0^2V_2  }{V_1+V_2},V_1+V_2,\frac{T_1V_1 +T_2V_2  }{V_1+V_2} ), Obs_1 :: Obs_2,\max(t_1,t_2))\\
     &[\![Observe(P,idn)]\!]^{\rho}_t=\\
       &\quad let\, ((x_0,V,T),Obs,t_1)=[\![P]\!]^{\rho}_t\\
       &\quad let\, O=(x_0,idn,t_1)\\
      &\quad  ((x_0,V,T),Obs\cup O,t_1)\\
       &[\![ Equilibrate(P,t)]\!]^{\rho}_{t'}=\\
       &\quad let\, ((x_0,V,T),Obs,t_1)=[\![P]\!]^{\rho}_{t'}\\
      &\quad  ([\![(\mathcal{A},\mathcal{R}),x_0,V,T)]\!](H)(t),Obs,t_1+t), 
\end{align*} 
where $H\in \mathbb{R}_{\geq 0}$ is such that for any $Equilibrate(P,t)$, $[\![(\mathcal{A},\mathcal{R}),$ $x_0,V,T)]\!](H)(t)$ is well posed. If such $H$ does not exist, we say that $P$ is ill posed.
\end{mydef}
Observations are stored as a list of strings, each of which  memorizing the concentration of the species at the observation, the identificator of the observation, and the observation time. 
Note that the above syntax does not prevent the programmer to assign the same identifier to two distinct observations.
We further stress that often observations of the state of an experiment are not exact, but corrupted by sensing noise. 
For instance, this is what happens with noisy fluorescence measurements. This noise can be easily taken into account at a semantical level by sampling an observation from a distribution with additive noise, where the noise level depends on the particular measure technique or instrumentation. 
Finally, we can also extend the sample semantics to take into account noise in Dispense operations. 

\vspace{-1.em}
\section{Case Study}\label{Sec-CaseStudy}
\begin{figure}
    \centering
    \includegraphics[width=0.8\textwidth]{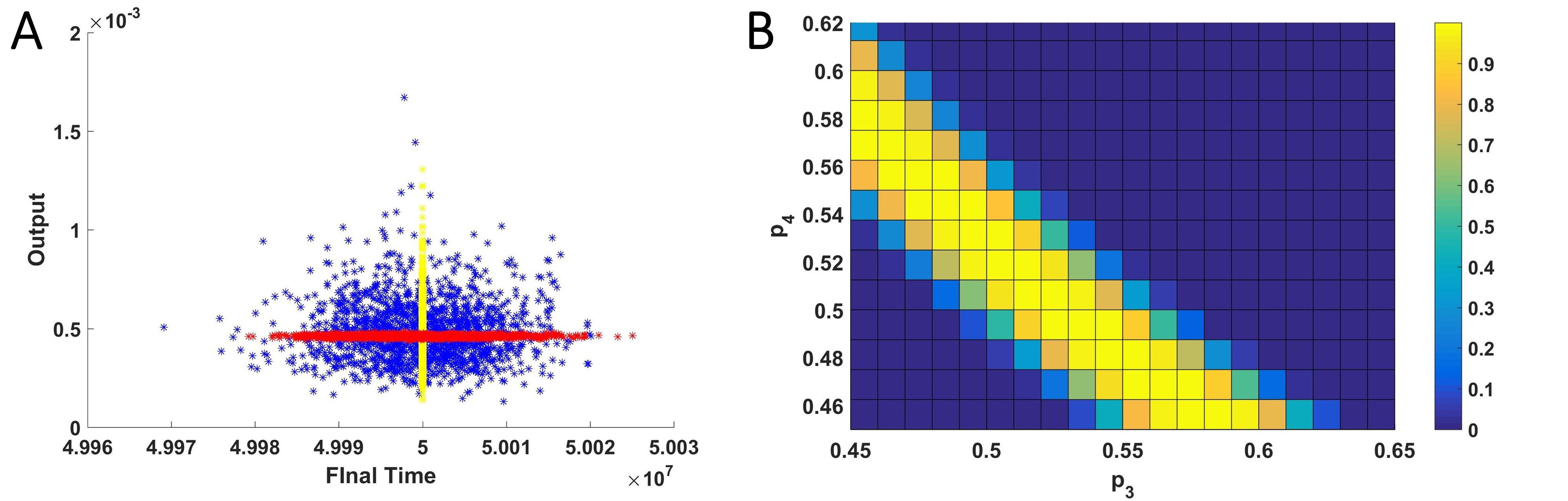}
    \caption{{(A): (red) $1500$ execution of the protocol assuming the physical model is fully known, and the only source of noise is in the discrete parameters of the protocols ($p_1,p_2,p_3,p_4$). (yellow) $1500$ executions of the protocol when the rates of the physical system are sampled from a sub-Poisson distribution, and discrete operations are not affected by noise. (blue) $1500$ simulations of the protocol when both sources of noise are active.(B):  $P_{Safe}([3.0\cdot10^{-4},3.5\cdot 10^{-4}])$ as a function of $p_3$ and $p_4$. Each cell is estimated from $20000$ executions of the protocol. }\vspace{-1.5em}}
    \label{fig:SampleOutput}
\end{figure}

\vspace{-0.5em}
As a case study we consider the experimental protocol for DNA strand displacement presented in Figure \ref{fig:DSDExampleFigure}. 
The protocol in Figure \ref{fig:DSDExampleFigure}a can be written formally as follows. We use $let\, x,_ = Dispense(P1, p)\, in\, P2$ as a short-hand for $let\, x,y = Dispense(P1,p)\, in\, Mix(Dispose(y),x)$
\begin{align*}
P_1=    &let\, In1 = ((Input1, 100.0 nM), 0.1 mL, 298.15 K)\, in\\
&let\, In2 = ((Input2, 100.0 nM), 0.1 mL, 298.15 K)\, in\\
&let\, GA = ((Output, 100.0 nM), 0.1 mL, 298.15 K)\, in\\
&let\, GB = ((Gate_B, 100.0 nM), 0.1 mL, 298.15 K)\, in\\
&let\, sGA, _ = Dispense(GA, p_1)\, in\\
&let\, sGB, _ = Dispense(GB, p_2) \,in\\
&let\, sIn1, _ = Dispense(In1,p_3)\, in\\
&let\, sIn2, _ = Dispense(In1,p_4)\, in\\
&Observe(Equilibrate(Mix(Mix(Equilibrate(Mix(sGA, sGB), t_1), sIn1), sIn2), t_2),idn),\vspace{-1em}
\end{align*}
where Input1, Input2, Output, $Gate_B$ are species of the CRN represented graphically in Figure \ref{fig:DSDExampleFigure}, $t_1=3000$, and $t_2=5 \cdot 10^6$.
According to the standard ISO 8655 for a volume of $1 mL$, 
the maximum standard deviation of a particular pipetting device is $0.3 \mu L$ per single operation. 
In order to incorporate such  an error in our model, we make use of the stochastic semantics. 
Thus, the concentration of the $Output$ strand at the end of the protocol is a random variable. 
It is also common that the reaction rates of the physical system are not known exactly and they may be affected by extrinsic noise \cite{paulsson2004summing}. This leads to another source of uncertanity in the output of the protocol, which can be easily incorporated in our semantics. We assume that the rate of each reaction has a normal distribution with variance equal to half of its mean (sub-Poisson noise). In Figure  \ref{fig:SampleOutput}a we plot $4500$ executions resulting from the protocol. From the figure it is easy to realize how the two difference sources of noise may  have a distinctive effect on the final outcome of the experiments. 

In many experimental protocols, 
one of the key challenges is to synthesize the optimal discrete parameters  to maximize the probability of obtaining desired behaviours. 
From now on, we assume perfect knowledge of the reaction rates of the physical system, while the discrete operations of the protocol and the times in each equilibration operation are still noisy. We assume $p_1=p_2=0.4$, and our goal is to see how the concentration of the $Output$ changes while varying $(p_3,p_4)\in [0.45,0.65]\times[0.45,0.65].$ We are interested in the following property
$$ P_{Safe}([3.0\cdot10^{-4},3.5\cdot 10^{-4}])=Prob(Output(t')\in [3.0\cdot10^{-4},3.5\cdot 10^{-4}]|t'=t_{final} ), $$
where $t_{final}$ is the final time of the protocol.
The following probability is estimated using Statistical Model Checking \cite{legay2010statistical} in Figure \ref{fig:SampleOutput}b,  
{ which in this context reduces to Monte-Carlo sampling.} 
 From Figure \ref{fig:SampleOutput}b it is easy to infer that the optimal value for such property is not unique (it is attained at values over the yellow band) and obtained, for instance, at  $(p_3,p_4)= (0.5,0.54)$.  
\vspace{-1.3em}
\section{Discussion}
\vspace{-0.7em}
We have presented a language to formalize experimental biological protocols, and provided both a deterministic and a stochastic semantics to this language. 
 Our language provides a unified description of the system being experimented on, together with the discrete events representing parts of biological protocols dealing with the handling of samples. Moreover, we allow the modeller to take into account uncertainties in both the  model structure and the equipment tolerances. This makes our language a suitable tool for both experimental and computational biologists.
 Our objective has been that of providing a basic language with an integrated representation of an experimental biological protocol. To this end, we have kept the language as simple as possible, showing how different extensions can be easily  integrated. 
For instance, in our denotational semantics, the dynamics of a physical process is given by a set of ODEs. This is accurate when the number of involved molecules is large enough, as in the discussed example of DNA strand displacement (DSD). However, in other scenarios, such as localized computation or gene expression, this might be unsatisfactory, as stochasticity becomes relevant \cite{arkin1998stochastic,dunn2015guiding}: the semantics presented here can be easily extended to incorporate such stochasticity, which can be achieved by considering more general classes of stochastic hybrid processes, such as switching diffusions \cite{laurenti2017reachability,cA10,yin2010hybrid} or continuous-time Markov chains (CTMCs) \cite{abate2015adaptive}. 

One of the main advantages in providing a language with formal semantics for experimental protocols is that protocols can now be quantitatively analyzed inexpensively in-silico, and classical problems of analysis of CRNs, such as parameter estimation \cite{cardelli2017syntax}, can be studied within the corresponding modelling framework; this can also take into account the discrete operations of the protocol, which influence the dynamics of the system. 
An additional target of this work is to provide automated techniques to synthesise optimal protocols, 
or to certify that protocols perform as desired. 
This can be attained by tapping into the mature literature on formal verification and strategy synthesis of PDMPs, 
or that of other more specialised models that the given protocol can be mapped onto. 
Notions of finite-state abstractions \cite{ZA14} and of probabilistic bisimulations \cite{cA10,abate2015adaptive},
as well as algorithms for probabilistic model checking of stochastic hybrid models \cite{laurenti2017reachability} will be relevant towards this goal. 






\bibliographystyle{abbrv}

{\tiny

\bibliography{LanguagesForBiology}
}

\section{Appendix}

We define the operation of substitution of a protocol into a variable as follows.
\begin{mydef}{(Substitution)}\label{DefinitionSubstitution}
$P_1\{x\leftarrow P_2\}$ is defined inductively as follows:  
\begin{align*}
&x\{x\leftarrow P \}=P\\ 
&y\{x\leftarrow P \}=y, \text{ for $x\neq y$}\\
&Mix(P_1,P_2)\{x\leftarrow P_3\} = Mix(P_1\{x\leftarrow P_3\},P_2\{x\leftarrow P_3\})\\
&(let\, x = P_1\, in \,P_2)\{x\leftarrow P_3\}  =  (let x = P_1\{x\leftarrow P_3\} \,in\, P_2)\\    
&(let\, y = P_1\, in\, P_2)\{x\leftarrow P_3\}  =  (let\, y = P_1\{x\leftarrow P_3\} in P_2\{x\leftarrow P_3\}),\\ 
&   \text{ for $x\neq y$ and $y \not \in FV(P_3)$} \\
&Equilibrate(P,t)\{x\leftarrow P_1\}=Equilibrate(P\{x\leftarrow P_1,t \})\\
&Dispose(P)\{x\leftarrow P_1\}=Dispose(P\{x\leftarrow P_1\}).
\end{align*}
\end{mydef}

\begin{mydef}{(Free Variables)}\label{Def-FreeVariables}
The set of Free Variables (FV) of a protocol $P$ is defined inductively as follows: 
\begin{align*}
&FV(x)=\{x\}  \\ 
&FV((x_0,V,T))=\{\}\\
&FV(Mix(P_1,P_2))=FV(P_1)\cup FV(P_2)\\
&FV (let\, x=P_1 \, in \, P_2)= FV(P_1)\cup(FV(P_2)-\{x\})\\
&FV (let \, x,y=Dispense(P_1,p)\, in\, P_2)= FV(P_1)\cup(FV(P_2)-\{x,y\})\\
&FV(Equilibrate(P,t))=FV(P)\\
&FV(Dispose(P))=FV(P).
\end{align*}
\end{mydef}

\end{document}